\documentclass[a4paper,11pt]{article}

\usepackage{jheppub}
\usepackage{tikz-cd}
\usepackage{adjustbox}
\usepackage{subcaption}
\usepackage{orcidlink}

\newcommand{\la}{\langle}
\newcommand{\ra}{\rangle}

\title{\Large Split-Helicity Tree Amplitudes and Flag Cluster Algebras}

\author[a, \orcidlink{0000-0003-2613-718X}]{Shruti Paranjape,}
\emailAdd{shruti\_paranjape@brown.edu}

\author[a, b, \orcidlink{0009-0005-6084-2466}]{Marcus Spradlin,}
\emailAdd{marcus\_spradlin@brown.edu}

\author[a, \orcidlink{0009-0008-2506-3207}]{Anastasia Volovich,}
\emailAdd{anastasia\_volovich@brown.edu}

\author[a, \orcidlink{0000-0003-3781-6153}]{He-Chen Weng}
\emailAdd{he-chen\_weng@brown.edu}

\affiliation[a]{\footnotesize Department of Physics,
    Brown University,
    Providence,
    RI 02912,
    USA
}

\affiliation[b]{\footnotesize Brown Center for Theoretical Physics and Innovation,
    Brown University,
    Providence,
    RI 02912,
    USA
}

\abstract{Recent work has uncovered a connection between the symbol letters of general massless scattering and (permutations of) cluster variables of partial flag varieties.  In this paper we explore the cluster adjacency of tree-level gluon amplitudes, specifically focusing on split-helicity amplitudes which can be written in closed form in terms of zigzag diagrams.  We check in several cases, and conjecture in general, that the poles in each term satisfy cluster adjacency under a set of permutations that is built from arc permutations of the corresponding zigzag.}

\begin{document}
\maketitle
\flushbottom

\section{Introduction}

Cluster algebras have been found to play a remarkable role in connecting the physical and mathematical properties of
scattering amplitudes.  In planar $\mathcal{N}=4$ SYM theory, the positions of poles and branch points of
amplitudes, captured by their symbol letters~\cite{Goncharov:2010jf}, are related to cluster variables~\cite{Fomin2001ClusterAI, gekhtman2002cluster, williams2014cluster} of Grassmannian cluster algebras~\cite{Scott2003GrassmanniansAC, Golden:2013xva}. Another important aspect of this connection is cluster adjacency, which ties physical analytic properties of amplitudes to the mathematical notion of cluster compatibility. At loop-level, cluster adjacency~\cite{Drummond:2017ssj} is the observation that the latter dictates which pairs of symbol letters are allowed to appear next to each other in an amplitude.
At tree-level, cluster adjacency~\cite{Drummond:2018dfd,Mago:2019waa} is the observation that the set of poles of every BCFW term in an amplitude are cluster variables from a common cluster; this has recently been proven rigorously~\cite{Even-Zohar:2023del} to be a property of BCFW tilings of the amplituhedron.

Most of the discussion regarding the cluster structure of $n$-particle amplitudes has utilized the close connection between the Grassmannian ${\rm Gr}(4,n)$ and the momentum twistor parameterization of kinematic space~\cite{Hodges:2009hk}.
Recently it has been appreciated that there is also a natural cluster structure~\cite{Bossinger:2024apm}
associated to the partial flag variety $\mathcal{F\ell}_{2,n-2;n}$ that describes the kinematic space of
massless $n$-particle scattering in terms of
traditional spinor helicity variables~\cite{Maazouz:2024qmm}.
It was recently shown that a large subset of the full set of planar massless two-loop 6-particle symbol
letters are indeed related to cluster variables of this algebra~\cite{Pokraka:2025ali, Bossinger:2025rhf}, closely mirroring the connection familiar in momentum twistor space for $\mathcal{N}=4$ SYM. One potentially
unfamiliar aspect of this construction, already pointed out
in an analogous construction~\cite{Bossinger:2022eiy} for the 5-particle case, is that it is necessary to consider allowing the kinematic variables to be embedded in the partial flag variety under arbitrary permutations.

Inspired by these developments, in this paper we explore whether individual terms of tree-level $n$-particle amplitudes expressed in spinor helicity variables exhibit cluster adjacency (under suitable permutations)
with respect to the cluster algebra associated to the partial flag variety $\mathcal{F\ell}_{2,n-2;n}$.
To begin with we focus on split-helicity amplitudes, which are particularly simple since they form a closed subset of helicity amplitudes under BCFW recursion and can all be obtained by successive gluing of three-particle MHV and $\overline{\rm MHV}$ amplitudes.  The particularly simple structure of these amplitudes allow them to be written down in closed form in terms of certain `zigzag diagrams'~\cite{Britto:2005dg}.  We use the Sklyanin bracket~\cite{Golden:2019kks} to test the cluster adjacency of the poles associated to individual zigzag diagrams. Based on extensive experimentation for $n \le 11$, we describe a specific set of permutations (which we call `good
permutations') associated to each zigzag diagram, and conjecture that these are precisely the permutations
under which the associated term satisfies (tree-level) cluster adjacency with respect to the
$\mathcal{F\ell}_{2,n-2;n}$ cluster algebra. A key role is played in our construction by certain
types of permutations known in combinatorics as arc permutations~\cite{Adin2010}.

The paper is organized as follows. In Sections~\ref{sec:partial flag} and~\ref{sec:sklyanin} we review partial flag cluster algebras and the Sklyanin bracket test for checking cluster adjacency. In Section~\ref{sec:goodperm} we describe how to associate a set of good permutations to each zigzag diagram. In Section~\ref{sec:four} we summarize our findings on testing the cluster adjacency of terms associated to individual zigzag diagrams, and we
end in Section~\ref{sec:discussion} with a discussion and some open problems.

\section{Review}
\label{sec:review}

In this section we review the cluster structure of spinor helicity brackets and Sklyanin bracket test for cluster adjacency.

\subsection{Partial flag cluster algebras}
\label{sec:partial flag}

For $n \ge 5$, the $n$-point spinor helicity variety is isomorphic to the partial flag variety $\mathcal{F\ell}_{2,n-2;n}$~\cite{Maazouz:2024qmm}. 
More explicitly, a point in $\mathcal{F\ell}_{2,n-2;n}$ is a pair of nested vector spaces $V_1\subset V_2\subset \mathbb{C}^n$ where the two vector spaces are 2 and $(n-2)$-dimensional. We can naturally represent the point by a $(n-2)\times n$ matrix $M$ where the vector spaces are the row span of the first 2 and $n-2$ rows. 
We also define Pl\"ucker coordinates for $\mathcal{F\ell}_{2,n-2;n}$ as the $2\times2$ and $(n-2)\times (n-2)$ minors of $M$
\begin{equation}
    P_{i_1i_2} = \text{det}(M_{a,b})_{a\in \{1,2\},\,b\in \{i_1,i_2\}}\quad P_{i_1\dots i_{n-2}}= \text{det}(M_{a,b})_{a\in \{1,\dots,n-2\},\,b\in \{i_1,\dots, i_{n-2}\}}\,.
\end{equation}
The isomorphism between spinor helicity brackets and the Pl\"ucker coordinates of $\mathcal{F\ell}_{2,n-2;n}$ is then simply
\begin{equation}
    \langle ij\rangle \leftrightarrow P_{ij} \quad\text{and}\quad [ij] \leftrightarrow (-1)^{1+i+j}P_{\{1,\dots,n\}\setminus \{i,j\}}\,,
    \label{eq: Spinor to flag}
\end{equation}
for $i<j$. 

\begin{figure}
     \centering
\adjustbox{scale=0.7}{\begin{tikzcd}
 	\boxed{\la 45 \ra} & {\boxed{\la15\ra}} & {\boxed{\la12\ra}} \\
 	\boxed{[12]} & {[23]} & {[34]} \\
 	\boxed{-[15]} & {[25]} & {-[35]} \\
     & & {\boxed{[45]}}\\
        \arrow[from=2-1, to=2-2]
        \arrow[from=2-2, to=2-3]
        \arrow[from=3-1, to=3-2]
        \arrow[from=3-2, to=3-3]
        \arrow[from=1-2, to=2-2]
        \arrow[from=2-2, to=3-2]
        \arrow[from=1-3, to=2-3]
        \arrow[from=2-3, to=3-3]
        \arrow[from=2-2, to=1-1]
        \arrow[from=2-3, to=1-2]
        \arrow[from=3-2, to=2-1]
        \arrow[from=3-3, to=2-2]
        \arrow[from=3-3, to=4-3]
\end{tikzcd}}
\quad\quad
\adjustbox{scale=0.7}{\begin{tikzcd}
 	\boxed{P_{45}} & {\boxed{P_{15}}} & {\boxed{P_{12}}} \\
 	\boxed{P_{345}} & {P_{145}} & {P_{125}} \\
 	\boxed{P_{234}} & {P_{134}} & {P_{124}} \\  & & {\boxed{P_{1,2,3}}}\\
        \arrow[from=2-1, to=2-2]
        \arrow[from=2-2, to=2-3]
        \arrow[from=3-1, to=3-2]
        \arrow[from=3-2, to=3-3]
        \arrow[from=1-2, to=2-2]
        \arrow[from=2-2, to=3-2]
        \arrow[from=1-3, to=2-3]
        \arrow[from=2-3, to=3-3]
        \arrow[from=2-2, to=1-1]
        \arrow[from=2-3, to=1-2]
        \arrow[from=3-2, to=2-1]
        \arrow[from=3-3, to=2-2]
        \arrow[from=3-3, to=4-3]
\end{tikzcd}}
\quad\quad
\adjustbox{scale=0.7}{\begin{tikzcd}
 	\boxed{\la 456 \ra} & {\boxed{\la 156 \ra}} & {\boxed{\la 126 \ra}} \\
 	\boxed{\la 345 \ra} & {\la 145 \ra} & {\la 125 \ra} \\
 	\boxed{\la 234 \ra} & {\la 134 \ra} & {\la 124 \ra} \\  & & {\boxed{\la 123 \ra}}\\
        \arrow[from=2-1, to=2-2]
        \arrow[from=2-2, to=2-3]
        \arrow[from=3-1, to=3-2]
        \arrow[from=3-2, to=3-3]
        \arrow[from=1-2, to=2-2]
        \arrow[from=2-2, to=3-2]
        \arrow[from=1-3, to=2-3]
        \arrow[from=2-3, to=3-3]
        \arrow[from=2-2, to=1-1]
        \arrow[from=2-3, to=1-2]
        \arrow[from=3-2, to=2-1]
        \arrow[from=3-3, to=2-2]
        \arrow[from=3-3, to=4-3]
\end{tikzcd}}
\caption{From left to right: the initial cluster of 5-point spinor helicity variety, the initial cluster of $\mathcal{F\ell}_{2,3;5}$ and the embedding of $\mathcal{F\ell}_{2,3;5}$ into $\text{Gr}(3,6)$, with frozen vertices boxed. 
In this example, the cluster algebra of $\mathcal{F\ell}_{2,3;5}$ is isomorphic to that of $\text{Gr}(3,6)$, but this is not the case in general.}
\label{figure: initial cluster 235}
\end{figure}
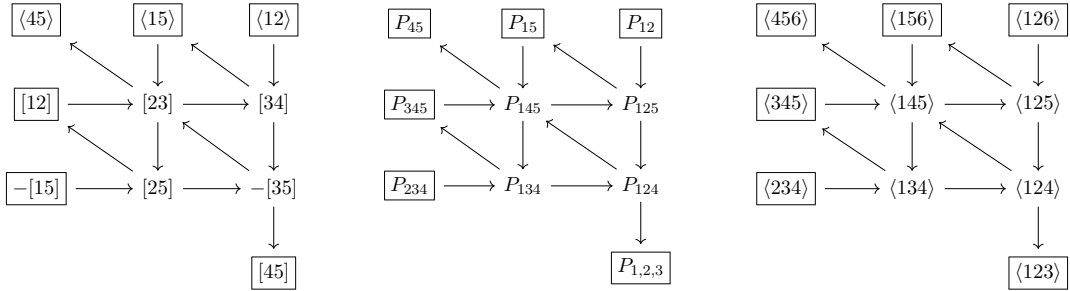

The cluster structure of the partial flag variety was first discussed in~\cite{Geiss2006PartialFV}, and it was later pointed out in~\cite{Bossinger:2024apm} that its cluster algebra admits an embedding into a Grassmannian cluster algebra\footnote{For an introduction to cluster algebras, see~\cite{williams2014cluster}.}. In particular, the $\mathcal{F\ell}_{2,n-2;n}$ cluster algebra is embedded into the $\text{Gr}(n-2,2n-4)$ cluster algebra through the map
\begin{equation}
    P_{i_1i_2} \to \langle i_1 i_2 (n+1)\cdots (2n-4)\rangle \quad \text{and}\quad P_{i_1\cdots i_{n-2}} \to \langle i_1\cdots i_{n-2}\rangle
    \label{eq: Flag to Grassmannian}
\end{equation}
where we denote the Pl\"ucker coordinates of $\text{Gr}(n-2,2n-4)$ with $\langle\cdots\rangle$. In Figure \ref{figure: initial cluster 235} we illustrate how the initial cluster of the 5-point spinor helicity variety (isomorphic to $\mathcal{F\ell}_{2,3;5}$) can be embedded into Gr(3,6). 

\subsection{Cluster adjacency test}
\label{sec:sklyanin}

The Sklyanin bracket~\cite{Golden:2019kks,Mago:2019waa} is conjectured to detect cluster adjacency for $\text{Gr}(k,n)$, i.e. whether two $\text{Gr}(k,n)$ variables share a mutual cluster or not. In this paper, we will test adjacency of spinor helicity expressions by first embedding the spinor brackets into the $\text{Gr}(n-2,2n-4)$ Pl\"ucker coordinates and then evaluating the Sklyanin bracket to determine adjacency. 

Concretely, two variables $\phi$ and $\psi$ are adjacent if and only if the Sklyanin bracket evaluates to a half-integer value,
$\{\log \phi,\log\psi\}\in \frac{1}{2}\mathbb{Z}$.
The Sklyanin bracket for $\text{Gr}(n-2,2n-4)$ can be easily computed as follows. We first use a suitable $\text{GL}(n-2,\mathbb{C})$ transformation, to gauge fix the $(n-2)\times(2n-4)$-dimensional matrix to the form
\begin{equation}
    \begin{pmatrix}
        1 & 0 & \cdots & 0 & y^1_1 & y^1_2 & \cdots & y^1_{n-2}\\
        0 & 0 & \cdots & 0 & y^2_1 & y^2_2 & \cdots & y^1_{n-2}\\
        \vdots & \vdots & \ddots & \vdots & \vdots & \vdots & \ddots & \vdots\\
        0 & 0 & \cdots & 1 & y^{n-2}_1 & y^{n-2}_2 & \cdots & y^{n-2}_{n-2}
    \end{pmatrix}.
\end{equation}
The Sklyanin bracket evaluated on the $y$ variables gives
\begin{equation}
    \{y^i_k,y^j_l\}=\frac{1}{2}(\text{sign}(j-i)-\text{sign}(l-k))y^i_l y^j_k\,.
\end{equation}
The action of the Sklyanin bracket on an arbitrary function of $y$ is then given by the chain rule
\begin{equation}
    \{f(y),g(y)\}=\sum_{i,j,k,l}^{n-2}\{y^i_k,y^j_l\} \frac{\partial f}{\partial y^i_k} \frac{\partial g}{\partial y^j_l}\,.
\end{equation}
In the following text, we interchangeably refer to a list of poles as `satisfying cluster adjacency' or `passing the Skylanin test' if the Sklyanin bracket evaluates to a half-integer value for every pair of elements in the list.

\section{Zigzag Diagrams and Good Permutations}
\label{sec:goodperm}

It was shown in~\cite{Britto:2005dg} that the tree-level split-helicity amplitude with $q$ negative helicity and $n{-}q$ positive helicity gluons can be expressed as a sum of $\binom{n-4}{q-2}$ terms labeled by `zigzag diagrams'.  In this section we explain how to associate to each zigzag diagram, i.e.~to each term in the amplitude, a certain collection of permutations (elements of $S_n$) that we call `good permutations'.

For $q>1$ and $n-q>1$ we draw a pair of parallel lines with points labeled $\{1,2,\ldots,q\}$ from left to right along the top line and $\{q{+}1,\ldots,n\}$ from right to left along the bottom line.  A zigzag is an oriented non-self-intersecting path consisting of an even number of line segments that begins between 1 and 2 on the top line and bounces between the lines (touching them at most once between each pair of neighboring labeled points) before ending on the top line between points $q{-}1$ and $q$.  (For the special case $q=2$ the zigzag is empty.)

To construct the good permutations associated to a zigzag diagram we proceed in four steps.
First, delete the segments shown here, if either exists:
\begin{align}
    \begin{tikzpicture}[scale=0.7, baseline={(0,-1.2cm)}]
        \draw[thick] (2,0) -- (-2,0);
        \draw[thick] (2,-3) -- (-2,-3);
        \draw[thick] (0,-0)-- (0,-3) node[pos=0.5, sloped] {$\blacktriangleright$};
        \node[draw=black, fill=white, minimum size=6pt] at (1,0) {};
        \node[draw=black, fill=white, minimum size=6pt] at (-1,0) {};
        \node[draw=black, fill=black, minimum size=6pt] at (1,-3) {};
        \node[draw=black, fill=black, minimum size=6pt] at (-1,-3) {};
        \node[above] at (-1,0.2) {1};
        \node[above] at (1,0.2) {2};
        \node[below] at (1,-3.2) {$n{-}1$};
        \node[below] at (-1,-3.2) {$n$};
    \end{tikzpicture}
    \qquad\qquad
    \begin{tikzpicture}[scale=0.7, baseline={(0,-1.2cm)}]
        \draw[thick] (2,0) -- (-2,0);
        \draw[thick] (2,-3) -- (-2,-3);
        \draw[thick] (0,-0)-- (0,-3) node[pos=0.5, sloped] {$\blacktriangleleft$};
        \node[draw=black, fill=white, minimum size=6pt] at (1,0) {};
        \node[draw=black, fill=white, minimum size=6pt] at (-1,0) {};
        \node[draw=black, fill=black, minimum size=6pt] at (1,-3) {};
        \node[draw=black, fill=black, minimum size=6pt] at (-1,-3) {};
        \node[above] at (-1,0.2) {$q{-}1$};
        \node[above] at (1,0.2) {$q$};
        \node[below] at (1,-3.2) {$q{+}1$};
        \node[below] at (-1,-3.2) {$q{+}2$};
    \end{tikzpicture}\qquad.
    \label{eq:eraserule}
\end{align}
In the special case $n = q+2$ this rule completely erases the zigzag; we discuss this and $q=2$ at the end of the section. Otherwise, the zigzag partitions the area between the two lines into a collection of $r > 1$ regions, two of which are unbounded.  We call the latter the `boundary regions' and we call the points in the bounded regions `internal points'.

Second, we mark precisely one point in each boundary region according to the rule
\begin{align}
\label{eq:markrule}
    \begin{tikzpicture}[scale=0.7, baseline={(0,-1.5cm)}]
        \draw[thick] (2,0) -- (-2,0);
        \draw[thick] (2,-3) -- (-2,-3);
        \draw[thick] (0,-0)-- (0,-3) node[pos=0.5, sloped] {$\blacktriangleright$};
        \node[draw=black, fill=white, minimum size=6pt] at (1,0) {};
        \node[draw=black, fill=white, minimum size=6pt] at (-1,0) {};
        \node[draw=black, fill=black, minimum size=6pt] at (1,-3) {};
        \node[draw=black, fill=black, minimum size=6pt] at (-1,-3) {};
        \node[above] at (-1,0.2) {$a$};
        \node[above] at (1,0.2) {$a{+}1^*$};
        \node[below] at (1,-3.2) {$b$};
        \node[below] at (-1,-3.2) {$b{+}1^*$};
    \end{tikzpicture}
    \qquad\qquad
    \begin{tikzpicture}[scale=0.7, baseline={(0,-1.5cm)}]
        \draw[thick] (2,0) -- (-2,0);
        \draw[thick] (2,-3) -- (-2,-3);
        \draw[thick] (0,-0)-- (0,-3) node[pos=0.5, sloped] {$\blacktriangleleft$};
        \node[draw=black, fill=white, minimum size=6pt] at (1,0) {};
        \node[draw=black, fill=white, minimum size=6pt] at (-1,0) {};
        \node[draw=black, fill=black, minimum size=6pt] at (1,-3) {};
        \node[draw=black, fill=black, minimum size=6pt] at (-1,-3) {};
        \node[above] at (-1,0.2) {$c^*$};
        \node[above] at (1,0.2) {$c{+}1$};
        \node[below] at (1,-3.2) {$d^*$};
        \node[below] at (-1,-3.2) {$d{+}1$};
    \end{tikzpicture}
\end{align}
applied to the edge on the boundary of that region.

Third, we assign a unique `target integer' to each internal point as follows.  Suppose there are respectively $n_R, n_L$  points in the right and left boundary regions.  Then, walking through the regions from right to left we assign target integers $\{n_R{+}1,\ldots,n{-}n_L\}$ to each labeled point in order, reading from right to left on both the top and bottom lines.

\begin{figure}
\begin{center}
    \begin{tikzpicture}[scale=0.7, baseline={(0,-1.5cm)}]
        \draw[thick] (17,0) -- (-2,0);
        \draw[thick] (17,-3) -- (-2,-3);
        \draw[thick] (1,-0)-- (6,-3) node[pos=0.5, sloped] {$\blacktriangleright$};
        \draw[thick] (6,-3)-- (5,0) node[pos=0.5, sloped] {$\blacktriangleleft$};
        \draw[thick] (5,0)-- (10,-3) node[pos=0.5, sloped] {$\blacktriangleright$};
        \draw[thick] (10,-3)-- (13,0) node[pos=0.5, sloped] {$\blacktriangleright$};
        \draw[thick] (13,0)-- (12,-3) node[pos=0.5, sloped] {$\blacktriangleleft$};
        \draw[thick] (12,-3)-- (15,0) node[pos=0.5, sloped] {$\blacktriangleright$};
        \node[draw=black, fill=white, minimum size=6pt] at (0,0) {};
        \node[draw=black, fill=white, minimum size=6pt] at (2,0) {};
        \node[draw=black, fill=white, minimum size=6pt] at (4,0) {};
        \node[draw=black, fill=white, minimum size=6pt] at (6,0) {};
        \node[draw=black, fill=white, minimum size=6pt] at (8,0) {};
        \node[draw=black, fill=white, minimum size=6pt] at (10,0) {};
        \node[draw=black, fill=white, minimum size=6pt] at (12,0) {};
        \node[draw=black, fill=white, minimum size=6pt] at (14,0) {};
        \node[draw=black, fill=white, minimum size=6pt] at (16,0) {};
        \node[draw=black, fill=black, minimum size=6pt] at (1,-3) {};
        \node[draw=black, fill=black, minimum size=6pt] at (3,-3) {};
        \node[draw=black, fill=black, minimum size=6pt] at (5,-3) {};
        \node[draw=black, fill=black, minimum size=6pt] at (7,-3) {};
        \node[draw=black, fill=black, minimum size=6pt] at (9,-3) {};
        \node[draw=black, fill=black, minimum size=6pt] at (11,-3) {};
        \node[draw=black, fill=black, minimum size=6pt] at (13,-3) {};
        \node[draw=black, fill=black, minimum size=6pt] at (15,-3) {};
        \node[above] at (0,0.2) {$1$};
        \node[above] at (2,0.2) {$2$};
        \node[above] at (4,0.2) {$3$};
        \node[above] at (6,0.2) {$4$};
        \node[above] at (8,0.2) {$5$};
        \node[above] at (10,0.2) {$6$};
        \node[above] at (12,0.2) {$7$};
        \node[above] at (14,0.2) {$8$};
        \node[above] at (16,0.2) {$9$};
        \node[above,text=red] at (2,1) {$13$};
        \node[above,text=red] at (4,1) {$12$};
        \node[above,text=red] at (6,1) {$9$};
        \node[above,text=red] at (8,1) {$8$};
        \node[above,text=red] at (10,1) {$7$};
        \node[above,text=red] at (12,1) {$6$};
        \node[above,text=red] at (14,1) {$4$};
        \node[below] at (1,-3.2) {$17$};
        \node[below] at (3,-3.2) {$16$};
        \node[below] at (5,-3.2) {$15^*$};
        \node[below] at (7,-3.2) {$14$};
        \node[below] at (9,-3.2) {$13$};
        \node[below] at (11,-3.2) {$12$};
        \node[below] at (13,-3.2) {$11^*$};
        \node[below] at (15,-3.2) {$10$};
        \node[below,text=red] at (11,-4.2) {$5$};
        \node[below,text=red] at (9,-4.2) {$10$};
        \node[below,text=red] at (7,-4.2) {$11$};
    \end{tikzpicture}
\caption{One of the 1716 zigzag diagrams that contribute to the 17-particle N${}^7$MHV amplitude. The initial labeled points are shown in black. The zigzag divides the area between the horizontal lines into 2 boundary and 5 internal regions. There are 10 internal points, $n_R=3$ points in the right boundary region, and $n_L=4$ points in the left. The rule~(\ref{eq:markrule}) marks boundary points 11 and 15.  The target integers $\{\color{red}4\color{black},\ldots,\color{red}13\color{black}\}$ are assigned uniquely to internal points as shown in red. There are $4 \times 8 = 32$ valid assignments of the remaining target integers $\{\color{red}1\color{black},\color{red}2\color{black},\color{red}3\color{black}\}$ and $\{\color{red}14\color{black},\color{red}15\color{black},\color{red}16\color{black},\color{red}17\color{black}\}$ to points in the right and left boundary regions respectively, and so a total of 64 `good permutations' associated to this zigzag diagram.}
\label{fig:bigexample}
\end{center}
\end{figure}
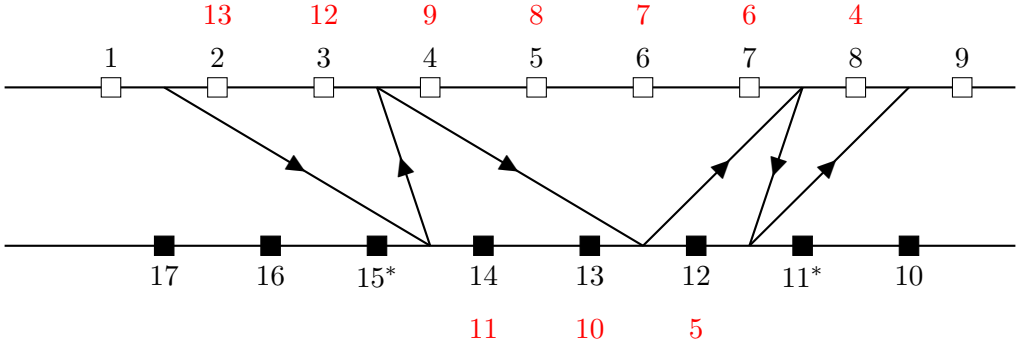

For example, for the zigzag diagram shown in Figure~\ref{fig:bigexample}, at this stage we have assigned the target integers
\begin{align}
\label{eq:example}
    \{1,2,\ldots,17\} \mapsto \{\cdot, 13, 12, 9, 8, 7, 6, 4, \cdot, \cdot, *, 5, 10, 11, *, \cdot, \cdot \}
\end{align}
to the internal points and it remains to describe the allowed ways to assign target integers $\{1,2,3\}$ to the right boundary points (in positions $\{9,10,11^*\}$) and target integers $\{14,15,16,17\}$ to the left boundary points (in positions $\{15^*,16,17,1\}$).

Finally, for each boundary region we start by choosing to assign either the smallest or largest possible target integer to the marked point. Then we assign to the neighbor of that marked point either the largest or smallest of the remaining target integers, and proceed until exhaustion. This process generates special cases of \emph{arc permutations}~\cite{Adin2010}.  We introduce a notation to denote the more restricted class of `rooted' arc permutations of the type that we require, using a star to denote the location of the root, which must be assigned either the largest or smallest integer. So for example we will write
\begin{align}
\begin{aligned}
{\rm Arc}(1,2,3,4^*) =\{ &\{1, 2, 3, 4\}, \{2, 1, 3, 4\}, \{2, 3, 1, 4\}, \{2, 3, 4, 1\}, \\
&\{3, 2, 1, 4\}, \{3, 2, 4, 1\}, \{3, 4, 2, 1\}, \{4, 3, 2, 1\}\}\,.
\end{aligned}
\end{align}
Altogether there are $2^{n_R-1} \times 2^{n_L-1}$ possible ways to assign target integers to the boundary points.
We summarize the 32 valid ways of completing~(\ref{eq:example}) by writing
\begin{align}
    \{17), 13, 12, 9, 8, 7, 6, 4, {\rm Arc}(1,2,3^*), 5, 10, 11, {\rm Arc}(14^*,15,16\}\,.
\end{align}

At this point in the procedure we have
described a set of $2^{n_R+n_L-2}$ elements of $S_n$ associated to each zigzag diagram.  We call these the `primary branch' permutations.  In addition, there is a secondary branch related by relabeling the target integers according to $i \leftrightarrow n{+}1{-}i$, obtained via the same procedure except reading through the zigzag diagram from left to right.
Together they comprise the set of $2^{n_R+n_L-1}$ `good permutations' we associate
to the zigzag diagram.

Let us comment on the edge cases of this construction where $q=2$, which can be thought of as an empty zigzag, or $q=n-2$, when the zigzag becomes empty after applying the rule~(\ref{eq:eraserule}).  These correspond respectively to MHV and $\overline{\text{MHV}}$ amplitudes.  In these cases the good permutations are those of the form $\{n), {\rm Arc}(1,2,\ldots\}$ or $\{ 2,\ldots,n), {\rm Arc}(1\}$.

\section{Cluster Adjacency of Split-Helicity Amplitudes}
\label{sec:four}

In this section we report our conjecture, based on numerous numerical checks, that each BFRSV term
in a split-helicity amplitude satisfies cluster adjacency with respect to any good permutation associated to the
corresponding zigzag diagram.

The main result of~\cite{Britto:2005dg} was the formula
\begin{align}
\label{eq:bfrsv}
A(1^-,\ldots,q^-,(q{+}1)^+,\ldots,n^+) =
\sum_{k=0}^{\min(q-3,n-q-2)} \sum_{A_k,B_{k+1}} \frac{N_1 N_2 N_3}{D_1 D_2
D_3}
\end{align}
where $A_k$ and $B_{k+1}$ respectively range over all subsets of
$\{2,\ldots,q{-}2\}$ and $\{q{+}1,\ldots,n{-}1\}$ of cardinality $k$ and $k{+}1$ respectively.
In increasing numerical order, their elements are labeled $a_1, a_2,\ldots,a_k$ and $b_{k+1}, b_k, \ldots, b_1$.
In terms of the zigzag diagrams reviewed in the previous section,
each term in this expression is associated to a zigzag with $2(k+1)$ edges that touches the
top line between the labeled points $1,2$, $a_1,a{+}1$, $a_2, a_2{+}1$, $\ldots$, $a_k,a_k{+}1$, $q{-}1,q$
and the bottom line between $b_1{+}1,b_1$, $b_2{+}1,b_2$, $\ldots$, $b_{k+1}{+}1, b_{k+1}$.
In terms of this data, the quantities appearing in the numerator and denominator are defined by
\begin{align}
\begin{aligned}
N_1 &= {\langle 1 | P_{2,b_1} P_{b_1 + 1, a_1} P_{a_1 + 1, b_2} \cdots
P_{b_{k+1} + 1,q - 1}|q\rangle^3}\,,\\
N_2 &=
\langle b_1{+}1~b_1 \rangle \langle b_2{+}1~b_2 \rangle
\cdots
\langle b_{k+1}{+}1~b_{k+1} \rangle\,,\\
N_3 &= [a_1~a_1{+}1] \cdots [a_k~a_k{+}1]\,,\\
D_1 &=
P^2_{2,b_1} P^2_{b_1 + 1, a_1} P^2_{a_1 + 1, b_2} \cdots
P^2_{b_{k+1} + 1,q - 1}\,,\\
D_2 &= F_{q,1} \overline{F}{}_{2,q-1}\,,\\
D_3 &= [2|P_{2,b_1}|b_1{+}1 \rangle \langle b_1|P_{b_1 + 1,a_1}|a_1]
[a_1{+}1|P_{a_1+1,b_2}|b_2{+}1\rangle
\cdots
\langle b_{k+1}|P_{b_{k+1}+1,q-1}|q{-}1]\,,
\end{aligned}
\end{align}
where  $P_{x,y} = p_x + p_{x+1} + \cdots + p_y$ and $F_{x,y}$ is given by
\begin{align}
F_{x,y} = \langle x~x{+}1\rangle \langle x{+}1~x{+}2\rangle
\cdots \langle y{-}1~y\rangle
\end{align}
with $\overline{F}_{x,y}$ given by the same expression but with the inner product
$[\cdot~\cdot]$.

We use the Sklyanin bracket test described in Section~\ref{sec:review} to check whether the denominator factors in each term of~(\ref{eq:bfrsv}) satisfy cluster adjacency with respect to any permutations of the partial flag cluster algebra $\mathcal{F\ell}_{2,n-2;n}$. For example, we find that the three terms that contribute to the 7-point split-helicity NMHV amplitude ($n=7, q=3$) satisfy cluster adjacency precisely with respect to the 64, 32, 64 corresponding good permutations (the dashed lines in the following denote edges removed by the rule~(\ref{eq:eraserule})):
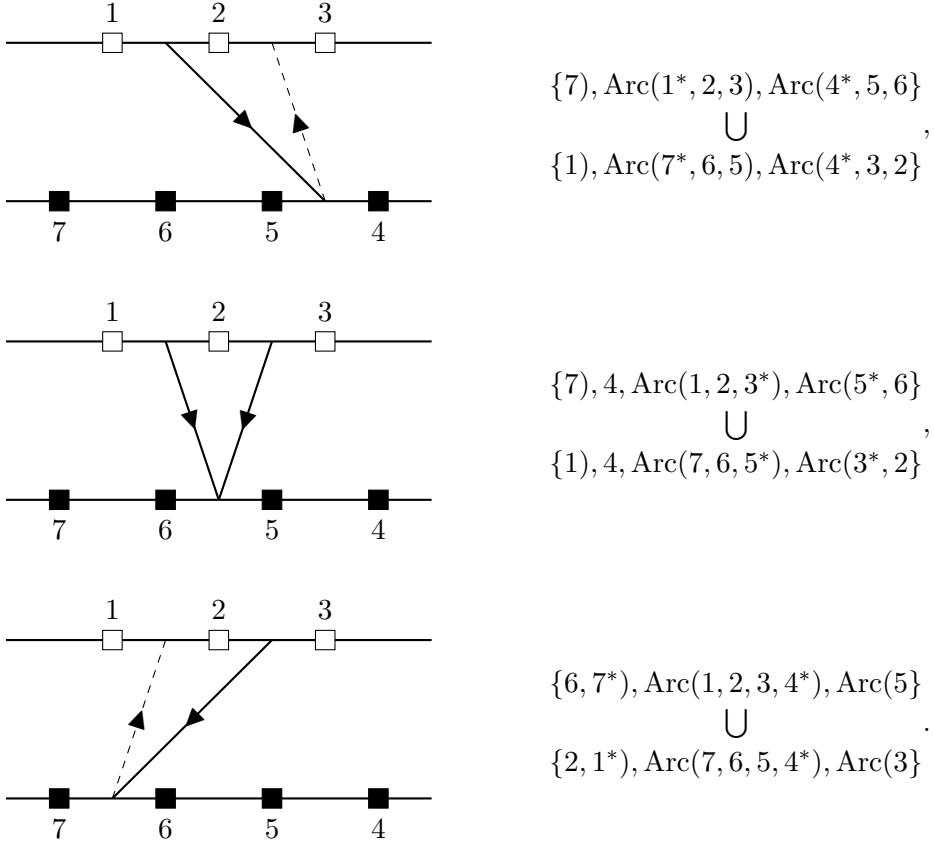
\begin{align*}
    \begin{tikzpicture}[scale=0.7, baseline={(0,-1.2cm)}]
        \draw[thick] (7,0) -- (-1,0);
        \draw[thick] (7,-3) -- (-1,-3);
        \draw[thick] (2,0)-- (5,-3) node[pos=0.5, sloped] {$\blacktriangleright$};
        \draw[dashed] (5,-3)-- (4,0) node[pos=0.5, sloped] {$\blacktriangleleft$};
        \node[draw=black, fill=white, minimum size=6pt] at (1,0) {};
        \node[draw=black, fill=white, minimum size=6pt] at (3,0) {};
        \node[draw=black, fill=white, minimum size=6pt] at (5,0) {};
        \node[draw=black, fill=black, minimum size=6pt] at (6,-3) {};
        \node[draw=black, fill=black, minimum size=6pt] at (4,-3) {};
        \node[draw=black, fill=black, minimum size=6pt] at (2,-3) {};
        \node[draw=black, fill=black, minimum size=6pt] at (0,-3) {};
        \node[above] at (1,0.2) {1};
        \node[above] at (3,0.2) {2};
        \node[above] at (5,0.2) {3};
        \node[below] at (0,-3.2) {7};
        \node[below] at (2,-3.2) {6};
        \node[below] at (4,-3.2) {5};
        \node[below] at (6,-3.2) {4};
    \end{tikzpicture}
    \qquad\qquad
    \begin{matrix}
    \{ 7), {\rm Arc}(1^*,2,3), {\rm Arc}(4^*,5,6\}\\
    \bigcup\\
    \{ 1), {\rm Arc}(7^*,6,5), {\rm Arc}(4^*,3,2\}
    \end{matrix},
\end{align*}
\begin{align*}
    \begin{tikzpicture}[scale=0.7, baseline={(0,-1.2cm)}]
        \draw[thick] (7,0) -- (-1,0);
        \draw[thick] (7,-3) -- (-1,-3);
        \draw[thick] (2,0)-- (3,-3) node[pos=0.5, sloped] {$\blacktriangleright$};
        \draw[thick] (3,-3)-- (4,0) node[pos=0.5, sloped] {$\blacktriangleleft$};
        \node[draw=black, fill=white, minimum size=6pt] at (1,0) {};
        \node[draw=black, fill=white, minimum size=6pt] at (3,0) {};
        \node[draw=black, fill=white, minimum size=6pt] at (5,0) {};
        \node[draw=black, fill=black, minimum size=6pt] at (6,-3) {};
        \node[draw=black, fill=black, minimum size=6pt] at (4,-3) {};
        \node[draw=black, fill=black, minimum size=6pt] at (2,-3) {};
        \node[draw=black, fill=black, minimum size=6pt] at (0,-3) {};
        \node[above] at (1,0.2) {1};
        \node[above] at (3,0.2) {2};
        \node[above] at (5,0.2) {3};
        \node[below] at (0,-3.2) {7};
        \node[below] at (2,-3.2) {6};
        \node[below] at (4,-3.2) {5};
        \node[below] at (6,-3.2) {4};
    \end{tikzpicture}
    \qquad\qquad
    \begin{matrix}
    \{ 7), 4, {\rm Arc}(1,2,3^*), {\rm Arc}(5^*,6\}\\
    \bigcup\\
    \{ 1), 4, {\rm Arc}(7,6,5^*), {\rm Arc}(3^*,2\}
    \end{matrix},
\end{align*}
\begin{align*}
    \begin{tikzpicture}[scale=0.7, baseline={(0,-1.2cm)}]
        \draw[thick] (7,0) -- (-1,0);
        \draw[thick] (7,-3) -- (-1,-3);
        \draw[dashed] (2,0)-- (1,-3) node[pos=0.5, sloped] {$\blacktriangleright$};
        \draw[thick] (1,-3)-- (4,0) node[pos=0.5, sloped] {$\blacktriangleleft$};
        \node[draw=black, fill=white, minimum size=6pt] at (1,0) {};
        \node[draw=black, fill=white, minimum size=6pt] at (3,0) {};
        \node[draw=black, fill=white, minimum size=6pt] at (5,0) {};
        \node[draw=black, fill=black, minimum size=6pt] at (6,-3) {};
        \node[draw=black, fill=black, minimum size=6pt] at (4,-3) {};
        \node[draw=black, fill=black, minimum size=6pt] at (2,-3) {};
        \node[draw=black, fill=black, minimum size=6pt] at (0,-3) {};
        \node[above] at (1,0.2) {1};
        \node[above] at (3,0.2) {2};
        \node[above] at (5,0.2) {3};
        \node[below] at (0,-3.2) {7};
        \node[below] at (2,-3.2) {6};
        \node[below] at (4,-3.2) {5};
        \node[below] at (6,-3.2) {4};
    \end{tikzpicture}
    \qquad\qquad
    \begin{matrix}
    \{ 6,7^*), {\rm Arc}(1,2,3,4^*), {\rm Arc}(5\}\\
    \bigcup\\
    \{ 2,1^*), {\rm Arc}(7,6,5,4^*), {\rm Arc}(3\}
    \end{matrix}.
\end{align*}

We have used the Sklyanin bracket test described in Section~\ref{sec:review} to check that each term in~(\ref{eq:bfrsv}) satisfies cluster adjacency with respect to the partial flag cluster algebra $\mathcal{F\ell}_{2,n-2;n}$ in the following cases and for all independent amplitudes (which means all values of $2 \le q \le \lfloor\frac{n}{2}\rfloor$).  For $n=6,7,8$ we have checked all permutations and verified that each term satisfies cluster adjacency only under the corresponding set of good permutations described in Section~\ref{sec:goodperm}.  For $n=9,10$ we verified that each term satisfies cluster adjacency with respect to all of the corresponding good permutations, and for $n=11$ we checked it for at least one good permutation for each term, though for $n>8$ we have not checked that cluster adjacency fails to hold for all of the `bad' permutations.

\section{Discussion}
\label{sec:discussion}

In this paper we have studied the cluster adjacency properties of individual terms contributing to split-helicity gluon amplitudes under permutations of the $\mathcal{F\ell}_{2,n-2;n}$ cluster algebra associated to $n$-particle spinor helicity variables.  Based on extensive experimentation, we conjecture that each term satisfies cluster adjacency precisely with respect to the corresponding `good permutations' described in Section~\ref{sec:goodperm}.  Our work leaves open several interesting questions.

The BFRSV terms in terms of which split-helicity amplitudes were expressed in~\cite{Britto:2005dg} are special cases of BCFW terms that can be used to express any helicity amplitude.  It is well-known that each of the latter is naturally associated (via a plabic graph, or equivalently on-shell diagram) to a unique permutation of $\{1,\ldots,n\}$~\cite{postnikov2006total,Arkani-Hamed:2012zlh}.  We find that the permutation associated in this way to each BFRSV term is not a `good' permutation in the sense of Section~\ref{sec:goodperm}, but given that each permutation determines an on-shell diagram and certain on-shell diagrams (those that correspond to contributions to split-helicity amplitudes) determine sets of good permutations, it would be interesting to skip the middleman and directly identify the combinatorial rule that determines the set of good permutations associated to (certain) individual permutations.

The momentum-twistor version of cluster adjacency for tree-level amplitudes, proven in~\cite{Even-Zohar:2023del}, is understood in the language of tiles of the amplituhedron~\cite{Arkani-Hamed:2013jha}, which describe contributions to the full tree-level superamplitude rather than individual gluonic component amplitudes. 
In~\cite{Even-Zohar:2023del}  the authors construct a cluster promotion map for the amplituhedron where the cluster adjacency of lower-point tiles extends to higher-point tiles. It is thus natural to ask if there exists a split-helicity version of the cluster promotion map that explains the cluster adjacency we find and perhaps explains the pattern of good permutations.

More generally, it is natural to wonder whether spinor-helicity cluster adjacency of the type we have studied could hold not just for split-helicity gluonic component amplitudes but for contributions to superamplitudes in non-chiral superspace~\cite{Huang:2011um} associated to tiles of the momentum amplituhedron~\cite{Damgaard:2019ztj}. 
Our initial investigation into this question has led to a negative result.  For example, the 7-point
NMHV superamplitude in non-chiral superspace is a sum of six terms (i.e.~there are six associated tiles),
only three of which contribute to the split-helicity gluon component amplitude.
We find that five out of the six terms do satisfy cluster adjacency (for at least one permutation each), but
for one term there is no permutation that makes cluster adjacency hold.

This seems to indicate some tension between the BCFW tiling of the momentum amplituhedron and the $\mathcal{F\ell}_{2,n-2;n}$ cluster structure. However, it is important to note that BCFW is one of many possible ways of tiling the amplituhedron (i.e.~of representing the superamplitude), so it is interesting to wonder whether there may be another representation that does exhibit cluster adjacency for each tile.

\acknowledgments

We would like to thank Matteo Parisi, Jian-Rong Li, Melissa Sherman-Bennett and Marcos Skowronek for valuable discussions. This work was supported in part by the US Department of Energy under contract DE-SC0010010 Task F and by Simons Investigator Award \#376208 (SP, AV). Part of this research was conducted using computational resources and services at the Center for Computation and Visualization, Brown University.

\bibliographystyle{JHEP}
\bibliography{biblio.bib}

\end{document}